\documentclass[10pt,journal,compsoc]{IEEEtran}
\pdfoutput=1

\ifCLASSOPTIONcompsoc
  \usepackage[nocompress]{cite}
\else
  \usepackage{cite}
\fi

\ifCLASSINFOpdf
\else
\fi

\usepackage{graphicx}
\begin{document}
\title{Gridlan: a Multi-purpose Local Grid Computing Framework}

\author{\'Attila~L.~Rodrigues and~Jo\~ao~Felipe~C.~L.~Costa%

\IEEEcompsocitemizethanks{\IEEEcompsocthanksitem A. Rodrigues and J. F. Costa are with Departamento de Engenharia de Minas of Universidade Federal do Rio Grande do Sul, Porto Alegre, Brazil.\protect \\

E-mail: attila.leaes$@$ufrgs.br, jfelipe$@$ufrgs.br.
}
} 

\IEEEtitleabstractindextext{%
\begin{abstract}
In scientific computing, more computational power generally implies  faster and possibly more detailed results. The goal of this study was to develop a framework to submit computational jobs to powerful workstations underused by nonintensive tasks. This is achieved by using a virtual machine in each of these workstations, where the computations are done. This group of virtual machines is  called the Gridlan. The  Gridlan framework is intermediate between the cluster and  grid computing paradigms. The Gridlan is able to profit from existing cluster software tools, such as resource managers like Torque, so a user with previous experience in cluster operation  can dispatch jobs seamlessly. A benchmark test of the Gridlan implementation shows the system's suitability for computational tasks, principally in embarrassingly parallel computations.
\end{abstract}
}

\maketitle

\IEEEdisplaynontitleabstractindextext

\IEEEpeerreviewmaketitle

\IEEEraisesectionheading{\section{Introduction}\label{sec:introduction}}

\IEEEPARstart{C}{omputer} simulation has became a major tool to solve many problems in science. However, the search for better results with higher precision, on problems of greater complexity, has led to a need for  increased computational power. One answer to this need is to use a collection of computers to perform  calculations. Such a collection is called a cluster of computers or a grid of computers, depending on its characteristics.

A traditional high-performance compu\-ting (HPC) computer cluster is characterized by  homogeneous hardware and software. A grid system is designed to solve the same types of problems as  clusters, but in a more complex environment in which  different computers may have different specifications and  network distances between  grid compu\-ting nodes may vary.

A well-known platform for grid computing is BOINC~\cite{boinc}. Developed from the SETI@Home~\cite{seti} project, this platform allows a researcher to run the required computations in the BOINC grid clients, which can be any PCs with  Internet connections. Other grid frameworks are XtremeWeb~\cite{xtreme}, Una\-Cloud~\cite{unacloud}, and OurGrid~\cite{ourgrid}. Each of these has a suite of tools to address  problems related to the characteristics of a grid environment.

A researcher who wants to use a grid platform such as BOINC must adapt their programs to run on the platform, a task that is generally considered difficult~\cite{grivas}.  Users of the grid client must also proactively choose projects to run, and any changes to a running project must be done interactively, i.e.,  grid client users must reconfigure the grid client software for each project.

Although this established grid system is appropriate for large projects where the increased computer power made available outweighs the disadvantages of having to spend considerable time and effort  dealing with the problems of grid implementation, there are numerous situations where researchers are faced with smaller scale computations on a daily basis  where such expenditure is not worthwhile.

The aim of our work is to develop a grid environment with the same procedures as an HPC cluster, with GNU/Linux as the operating system. The user must login with SSH into the grid server, compile the desired source code, send a job to an installed resource manager, and monitor the job with the usual HPC tools. 

Our specific grid environment (a local grid) is composed of computers connected to the grid's server LAN. It is targeted typically at computers in the same building as the designated server. Thus, our study concerns the development of a framework to aggregate computers in a situation intermediate  between grid computing and  cluster computing. Another guiding principle of the project is the need to run concurrently in a possible pre-existing cluster server. In this situation, there is an extra benefit, in that a user who wants to submit calculations may choose in the same server the resource manager's queue corresponding to the grid infrastructure or the cluster nodes. Another requirement imposed on the project is the need to remain invisible to the client. The installed software must not disrupt the usual work of  ordinary users of the machine or impose any specific operating system.

The project was carried out in our local research laboratory, using graduate students'  workstations. The computational power available from this setting cannot be ignored. The  machines have at least  quad core processors, with RAM ranging from $8$GB  to $32$GB. This  is not a controlled environment like a cluster network, but it is not geographically distributed as in the grid computing paradigm. This set of machines is herein after referred as the Gridlan. The objective is to treat each machine like a computer node in the Gridlan. The following sections explain how this was achieved.

The paper is structured as follows. Section~\ref{sec:framework} presents the proposed Gridlan framework and each relevant aspect of the project. Section~\ref{sec:implementation} describes the details of the implementation and software choices and gives benchmark results. Section~\ref{sec:best} discusses the best practices for using the Gridlan. Section~\ref{nextsteps} describes some issues that arise and discusses the next steps of this work. Section~\ref{sec:conclusion} presents our  conclusions.

\section{Gridlan framework}
\label{sec:framework}

\begin{figure}[!t]
\centering
\includegraphics[width=8cm]{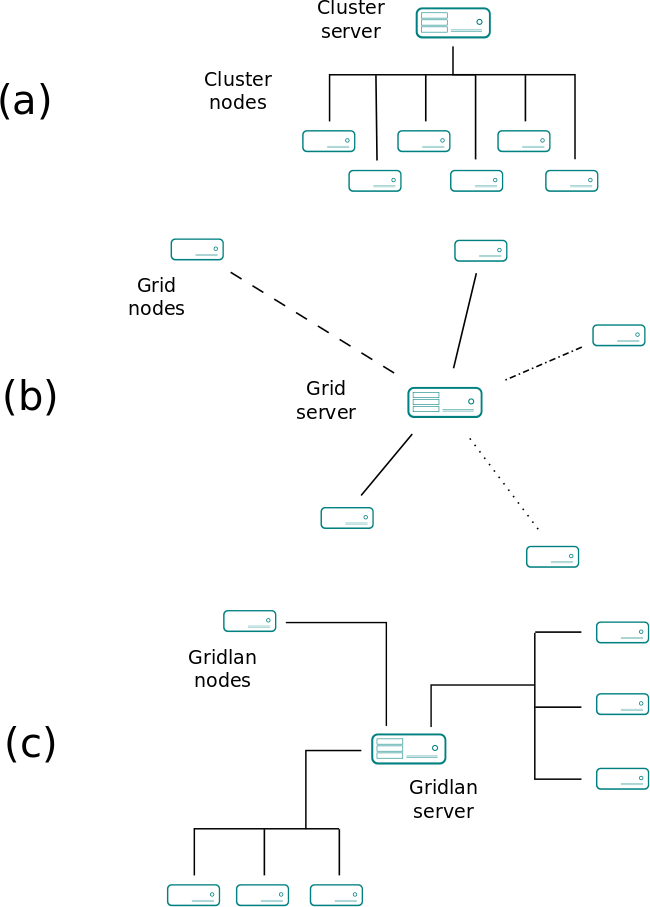}
\caption{Typical network schematics for clusters, grids, and the proposed Gridlan. (a) In a cluster infrastructure, the network distance is usually homogeneous. (b) A grid framework works with  heterogeneous connections.  (c) In the Gridlan proposal, the  network topology is intermediate between those of the cluster and the grid. Typically, the clients are a few switches or routers away from the server and are linked to it via wired connections.}
\label{schematic-network}
\end{figure}

The structure of the proposed computer network is intermediate between that of a cluster  and that of a grid (Fig.~\ref{schematic-network}). The Gridlan components are the Gridlan server and the Gridlan clients. The  clients are a heterogeneous pool of computers connected to the server.

The objective of the Gridlan project is to provide to the user the same experience as in a conventional  HPC cluster, including the ability to use the same tools that are available in such a cluster. The typical steps involved in submitting a computation to a cluster are as follows:

\begin{enumerate}
 \item The user connects to the  server using an SSH client.
 \item The user compiles the source code and prepares a script to submit the calculations to the resource manager.
 \item The user submits the job to the resource manager.
\end{enumerate}

There is some latitude in the choice of subsystems such as the resource manager. We used Torque~\cite{torque}, but others are also suitable.

\subsection{VPN connection from clients to server}

The first interaction between a client and the Gridlan server is a VPN connection. The client connects to the VPN server during the client's operating system start-up.

\begin{figure}[!t]
\centering
\includegraphics[width=9cm]{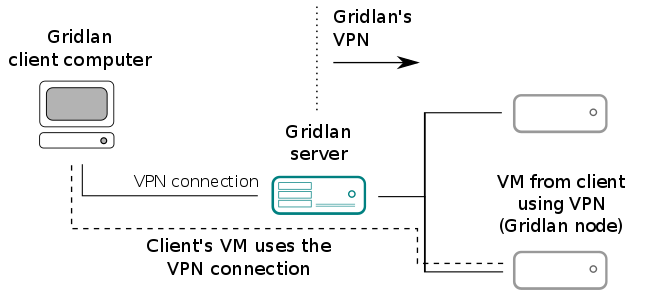}
\caption{A Gridlan client connected to the Gridlan VPN. The virtual machine will boot and communicate through the VPN connection.}
\label{schematic-network-vpn}
\end{figure}

With a VPN connection, the Gridlan client can interact with the necessary services hosted by the server as if it were in the same physical subnet (Fig.~\ref{schematic-network-vpn}). This approach facilitates the configuration of the necessary services  such as the network filesystem, the DHCP server, and the job manager system. To add a new client to the Gridlan VPN, a private key must be created by the server administrator and copied to the new client. This procedure provides the proper authorization allowing a computer to participate in the VPN.

Another characteristic of this setup is that the network traffic  is all routed via the Gridlan server. When two nodes exchange data, the latter always passes through the Gridlan server.

\subsection{Virtualized nodes}

To achieve the desired cluster experience in the Gridlan system, the researcher who wants to perform a calculation must not need to have any concerns about  operating system heterogeneity across the Gridlan clients. The solution is to run a virtual machine in these clients. With a virtual machine on each client, it is possible to  establish a homogeneous operating system to run  applications. Another advantage of this approach is that it isolates the host's operating system from the virtual machine where the grid's calculations take place. Virtualized nodes for computations are not a new feature of scientific computing infrastructure, and indeed are becoming increasingly popular in cloud computing~\cite{sandbox}~\cite{cloud1}~\cite{Huang}.

A virtual machine that is part of the Gridlan is referred to as a Gridlan node, or just  a node.

\subsection{Remote boot and nfsroot}\label{sec:nfsroot}

Any computing node needs an operating system and a filesystem to boot. In the Gridlan, the computing nodes are virtual machines started by the Gridlan clients. Usually, a virtual machine uses an ISO or disk image to boot, but, for system administration purposes, we chose a remote boot approach. The remote boot technique (also called diskless nodes) means that the node will request (using a DHCP request from the PXE) the necessary files from the Gridlan server to boot. After booting the Linux kernel and the initramfs, the virtual machine will mount the root filesystem via NFS. This technique is called nfsroot. All virtualized computing nodes share the same root filesystem. This exported filesystem resides at the Gridlan server, in the \verb|/nfsroot| folder.

The remote boot approach centralizes  maintenance in the server. To update a kernel, a new one must be compiled and copied to the TFTP directory (the TFTP service in the Gridlan server, which is responsible for sending the Linux kernel to the booting node). To install  new software in the nodea, the administrator must change the nodes' system in the folder \verb|/nfsroot|, with the command \verb|chroot /nfsroot apt-get install package|, for example. The \verb|/nfsroot| folder in the Gridlan server is the root filesystem for the Gridlan nodes.

\subsection{Procedure for sending jobs}

One of our guiding principles for this project concerns the job submission procedure. This must be as similar as possible  to the usual cluster submission procedure defined at the start of  Section~\ref{sec:framework}. 

A solution that keeps as close as possible to this procedure is to add the poll of Gridlan nodes as a new queue in the resource manager (Torque in our implementation). A special queue for the Gridlan nodes helps  users  choose the appropriate resources for their calculations.

The procedure to submit a job in Gridlan thus becomes the following:

\begin{enumerate}
 \item The user connects to the server using an SSH connection.
 \item The user chooses a queue to run the job and changes the Torque script accordingly.
 \item The user submits the job script to Torque.
 \item The queue manager finds the appropriate resources and runs the job.
\end{enumerate}
This new procedure contains one additional item, namely, the choice of the queue and the passage of the parameter that represents the selected queue in the submission process to the resource manager (Torque).

This procedure has a centralizing characteristic with regard to submissions. For a user to submit a job, there must first be a connection to the server. This ensures that the user has proper authorization to run the job on the grid. Even if a client connects to the grid, this  does not mean that users can run jobs on the Gridlan.

\subsection{Node initialization procedure}

Here, we summarize the client node initialization procedure, indicating where it is a virtual machine operation (node) and where it is a client computer operation. The process can be summarized as follows:

\begin{enumerate}
 \item The client (the virtual machine's host) connects to the cluster server using a VPN connection.
 \item The computer client starts the Gridlan node (the virtual machine).
 \item The virtual machine sends the DHCP requests through the VPN's tunnel, and all  communications between the node and the server are established via the VPN.
 \item The cluster server responds to the DHCP requests and sends the appropriate files for the node's initialization.
 \item The virtual machine mounts by NFS the filesystem root mount point ``/'' and finishes the operating system boot.
\end{enumerate}

\subsection{Node fault tolerance}

Another aspect of Gridlan development is the resilience of the nodes. The computer clients are unreliable, since they can be inadvertently turned off or can be victims of a network connection fault. Therefore, a simple procedure for checking  node connections was developed.

On the Gridlan server side, a script pings each node, saving the node state (on or off). This procedure is executed every 5 minutes. A script in the client machine asks the server if the virtual machine (the Gridlan node) is on. If the status is ``off,'' then a script to restart the node is executed.

\section{Gridlan implementation}
\label{sec:implementation}

\subsection{Hardware}

The Gridlan network comprised  computers at our university laboratory that were used by graduate students. The  heterogeneity of the hardware can be seen from Table~\ref{table_clients}. No changes were made to the computers' network connections in the laboratory, such as network cables and intermediate switches. Thus, the original computer configuration was used. 

\subsection{Choice of software components and possible alternatives}

The  Gridlan framework allows for different choices of software to perform desired functions. For example, the resource manager in our tests was  Torque 2.4.16; however, this could be changed to other software with the same function, such as  SLURM~\cite{slurm}, OpenLava~\cite{openl}, or HTCondor~\cite{condor}.

The main software  components used are as follows:

\begin{itemize}
\item Node and server operating system:  GNU/Linux (here Debian $8.5$~\cite{debian}, but any other Linux distribution could be used instead).
\item Transmission of the node kernel through the network: TFTP~\cite{tftp}. An alternative is iPxe~\cite{ipxe}, which  can be configured to use an HTTP connection.
\item Virtual machine software on hosts: QEMU~\cite{qemu}/KVM~\cite{kvm} in GNU/Linux;  VirtualBox~\cite{virtualbox} in Windows. An alternative is VMware~\cite{vmware}.
\item VPN connection: OpenVPN 2.3.4.
\item Node filesystem (nfsroot): NFS v3;
\item Resource manager: Torque 2.4.16~\cite{torque}. Alternatives are SLURM~\cite{slurm}, OpenLava~\cite{openl}, and HTCondor~\cite{condor}.
\end{itemize}

\begin{table}[!t]
\renewcommand{\arraystretch}{1.3}
\caption{Gridlan clients in the experiment. The total number of cores is $24$.}
\label{table_clients}
\centering
\begin{tabular}{c|c|c|c}
Node & Processor & No. of cores & Client OS \\
n01 & Xeon E5-2630  & $12$ &  GNU/Linux (Debian 8.1)  \\
n02 & Core i7-3930K & $6$ & Windows 10\\
n03 & Core i7-2920XM & $4$ & Windows 10\\
n04 & Core i7 960 & $4$ & Windows 7\\
\end{tabular}
\end{table}

\subsection{Network latency}

The latency between the nodes is expected to be larger than that between the Gridlan clients, since the data must travel via additional layers (the VPN connection and the virtual machine) to arrive at the final destination. To measure how much overhead in latency is introduced by the Gridlan framework, the usual ICMP ping was used, with $56$ bytes of payload. Two latency tests were carried out: one between the server and the clients (the host machine) and one between the server and the nodes (the virtual machine in each client). 
An additional latency test was also carried out: from the server to a client and from the server to the respective client's node with an MPI latency test using the same $56$ bytes for the message as the default ICMP ping. This additional test was intended to check whether the ping test was appropriate for the usual scientific computation tools like MPI.

The results for  the latency in node n01 are $1200(80)\,\mu$s for the MPI latency test and  $550(20)\,\mu$s for the ping test. These results are consistent with the ICMP ping results in Table~\ref{table_ping} for node n01.

The results for the measured latencies with the ICMP ping are shown in Table~\ref{table_ping}. The additional overhead provided by the Gridlan is roughly $900\,\mu$s.

The latency results show the expected overhead when data passes through the additional layers (VPN and virtual machine). This result indicates difficulties that may be encountered when running tightly interconnected computations across the Gridlan.

\begin{table}[!t]
\renewcommand{\arraystretch}{1.3}
\caption{Ping from Gridlan server.}
\label{table_ping}
\centering
\begin{tabular}{c|c|c}
Node & Client ping (host) & Node ping (VM)    \\
n01 & $550(20)\,\mu$s & $1250(30)\,\mu$s    \\
n02 & $660(20)\,\mu$s & $1500(110)\,\mu$s   \\
n03 & $750(40)\,\mu$s & $1650(90)\,\mu$s   \\
n04 & $610(30)\,\mu$s & $1400(100)\,\mu$s   \\
\end{tabular}
\end{table}

\subsection{Benchmarking  Gridlan computing power}

In order to provide data illustrating the full potential of the Gridlan, the grid performance was measured using independent parallel computations. The NPB Benchmark~\cite{npb} with EP kernel (Class D) was chosen for the tests.

For each run, a random number of Gridlan cores were chosen, from $1$ to $26$ (the latter being the total number of Gridlan cores in our implementation). The processes were then scattered randomly through the Gridlan clients, taking account of the number of available cores of each client. With this procedure, we can estimate the variation in  results caused by the heterogeneity of the client  hardware.

We performed the same NPB-EP class D test on another server to provide a comparison. The server used for this purpose had four AMD Opteron 6376 processors, giving a total of $64$ cores.

\begin{figure}[!t]
\centering
\includegraphics[width=9cm]{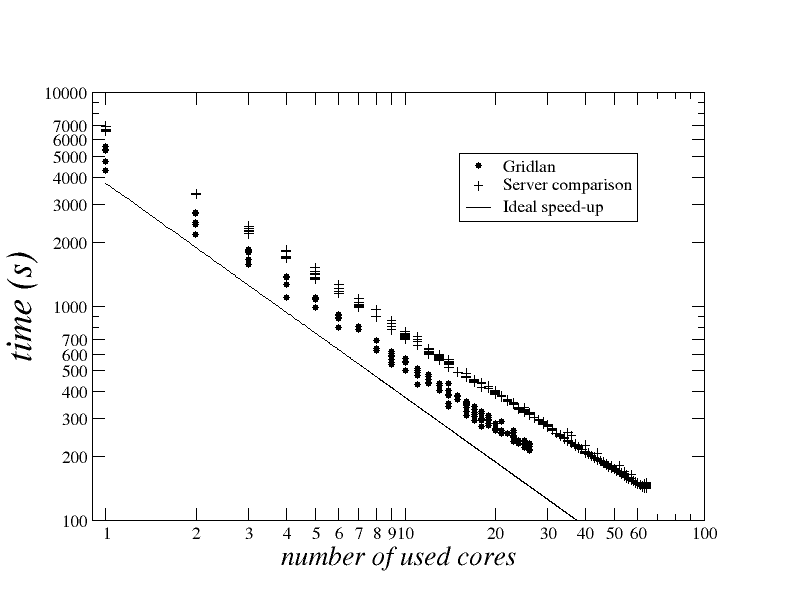}
\caption{Speed-up test. The results of the runs on the Gridlan are shown by filled circles and those on the comparison server by crosses. The continuous line represents the ideal speed-up $t(n)=t_1/n$, where $n$ is the number of cores used and $t_1$ is the time required to perform the calculations with one core. The results of the test do not agree with the ideal speed-up line owing to the Turbo Boost/Turbo Core technology used in the processors.}
\label{comparison}
\end{figure}

The benchmark results are shown in Fig.~\ref{comparison}, which displays the elapsed time of the benchmark test (NPB EP class D) for each run using the number of cores specified on the horizontal axis (an alternate form of Amdahl's law~\cite{amdahl}). It can be seen that the Gridlan group of four computers outperforms the comparison server for all tests up to the maximum number of Gridlan cores (26). The elapsed time for 26 Gridlan cores is about 212\,s; to achieve the same performance, the comparison server requires 38 cores.

As well as revealing that the Gridlan framework can   outperform the comparison server, the results also show the potential of this framework for numerical processing. Although  high network latency prohibits highly parallel computations, a user can perform this type of computation using one Gridlan node with a suitable number of cores.

It is worth noting that although the computations are embarrassingly parallel, the results (Fig.~\ref{comparison})  do not agree with the ideal speed-up defined by $t(n)=t_1/n$, where $n$ is the number of cores and $t_1$ is the time consumed by the calculations using one core. This phenomenon is due to the technology implemented in the processors whereby  the core's clocks are dynamically changed to increase performance when the number of intensive tasks is less than the number of cores  in the computer (this technology is called Turbo Boost by Intel and Turbo Core by AMD).

\section{Best practices for using the Gridlan}
\label{sec:best}

The heterogeneity of the Gridlan network  poses challenges with regard to  parallel computational performance. The best use of local grid resources is through independent simulations (also called embarrassingly parallel computations). In this case, each job submission corresponds to a process that will not interact with other processes during the calculation. For example,  this frequently happens in Monte Carlo calculations, where a statistical average of several simulations of the same experiment must be performed. Another example is when the goal of the calculation is to determine a curve from some simulation test, and each point of the curve is independently obtained from other points using different simulation parameters.

Simulations with intercommunicating processes can also be performed on the Gridlan. However, a careful analysis of the communication between the processes that comprise the simulation is needed. Usually, processes involved in parallel computation perform calculations at different times from those at which they communicate with each other or with a master process.

During the times at which processes communicate, the nodes do not perform any computational tasks, i.e., the allocated CPU is idle. If this communication time interval is much smaller than the effective computation time, then  parallelism is a suitable option. The limit of a job where its processes have  computation times much greater than their interconnection times is equivalent to a series of independent processes. An  example of an intermediate case  would be a process that spent 70\% of the time performing calculations and 30\% of the time communicating. It would be up to the user to decide whether this parallelization algorithm was acceptable for running on the Gridlan. It is important to note that this analysis should be performed regardless of the cluster structure used. In any cluster, there is a minimum latency between process data transfer, and the maximum bandwidth of data transmission has to be considered.

Another factor to take into account is the greater probability for one node of the Gridlan to go down compared with what happens in a conventional HPC cluster. This situation can arise in association with an accident (switching off a client inadvertently, for example) or with a system crash, or it can be caused by interruptions to the  electrical power supply  or by network events. One technique to improve the resilience of  submitted jobs is to write all the qsub scripts in a temporary folder. The last qsub script command must be to delete (or rename) the script. In this way, the unfinished job's scripts will still remain in the scripts folder and can be restarted later.

\section{Issues and next steps}
\label{nextsteps}

In Windows clients, a VirtualBox headless virtual machine is initialized as the operating system starts. For this confi\-gu\-ra\-tion,  the virtual machine instance must be run by the SYSTEM user. The consequence of this is that for a local user to run another VirtualBox instance, this must be done with administrator privileges. This  violates the guideline that states that the installed software must not interfere with  local users' experience. To address this issue, the VirtualBox instances can be replaced by another hypervisor such as pure QEMU, which can provide CPU emulation, although this is at the cost of a drop in  performance~\cite{vmtest}.

In the performance realm, optimizations in the VPN layer can be taken into account for  overhead latency mini\-mi\-zation between the server and the nodes~\cite{vpn}.

Another interesting feature that might be implemented in the future is a schedule for the clients.  Clients could then be tagged and the administrator could set a schedule specifying when jobs may be received from particular groups of clients. One example is a user who offers his computer for use by the local grid at  nighttime and  weekends. During daytime in this example,  unfinished jobs can be frozen and resumed later when the schedule permits. 

The installation process is currently not user-friendly owing to the nature of  many parts of the software confi\-gu\-ra\-tion and to some of the procedures involved, although these can be changed, as has been mentioned in the case of   VirtualBox in Windows clients. Therefore,  efforts to simplify and automate  software installation, principally on the client side, will be important further steps for the project. 

\section{Conclusion}
\label{sec:conclusion}

We have presented a technique for building an inexpensive computing cluster from computers that are being used for nonintensive tasks, the Gridlan. The Gridlan framework is aimed to fill the gap between the usual cluster HPC architecture and the grid computing paradigm,  relying on a VPN network of virtual machines where the calculations are performed.

The type of computations usually submitted to a grid are suitable to run on the Gridlan. However, jobs that normally run on a cluster need to be evaluated individually to determine whether they will run properly on the Gridlan.

The main characteristics of the Gridlan are its similarity to the usual cluster environment,  the ability of users to submit jobs via Torque (or another resource manager), and the facilitation of administration of  computing nodes via remote boot (PXE) and remote filesystem (nfsroot) resources.

Benchmark tests were realized on four computers in our laboratory, and a comparison with a local server showed that the performance of  the Gridlan  is such that it is appropriate for use in the following cases: independent computations (also called embarrassingly parallel computations), tightly integrated computations where the processes are in the same node, and parallel computations where the time spent making interconnections between  processes is negligible compared with the time spent performing calculations.

The relevant software, configuration details, and instructions to build a Gridlan can be found at the Gridlan page on GitHub~\cite{github}.

\section*{Acknowledgment}

This project is supported by Conselho Nacional de Desenvolvimento Cient\'\i fico e Tecnol\'ogico (CNPq).

\ifCLASSOPTIONcaptionsoff
  \newpage
\fi

\bibliographystyle{IEEEtran}
\bibliography{IEEEabrv,mybib.bib}

\end{document}